\documentclass[epjST]{svjour}

\usepackage{graphicx}
\usepackage{amssymb,amsmath,bm}
\usepackage{array}

\newcommand{\bff}[1]{\mbox{\boldmath ${#1}$}}

\newcommand\as{\ensuremath{\alpha_s}}
\newcommand\lambdaQCD{\ensuremath{\Lambda_{\rm QCD}}}
\newcommand\LambdaQCD{\lambdaQCD}
\newcommand\MSB{\ensuremath{\overline{\rm MS}}}

\newcommand\mB{\ensuremath{\overline{m}}}
\newcommand\mr{\ensuremath{m}}
\newcommand\mum{\ensuremath{\mu_m}}

\newcommand\mpole{\ensuremath{m}}

\newcommand\nl{n_l}
\newcommand\cf{C_F}
\newcommand\ac{c^{(\rm as)}}
\newcommand\acb{\tilde{c}^{(\rm as)}}

\begin{document}
\allowdisplaybreaks

\title{Pole mass renormalon and its ramifications}
\author{Martin Beneke\inst{1}
}
\institute{Physik Department T31, 
Technische Universit\"at M\"unchen, 
James-Franck-Stra\ss e~1,
D - 85748 Garching, Germany
}
\abstract{
I review the structure of the leading infrared renormalon divergence 
of the relation between the pole mass and the \MSB~mass of a heavy 
quark, with applications to the top, bottom and charm quark. That 
the pole quark mass definition must be abandoned in precision 
computations is a well-known consequence of the rapidly diverging 
series. The definitions and physics motivations of several 
leading renormalon-free, short-distance mass 
definitions suitable for processes 
involving nearly on-shell heavy quarks are discussed.
} 
\maketitle


\section{Introduction}
\label{intro}

The existence of a short-distance scale $Q\gg \Lambda_{\rm QCD}$ and 
infrared (IR) finiteness are important requirements for the application 
of perturbation expansions to strong interaction (QCD) processes, but they 
are not always enough in practice. The perturbative series is divergent, 
most likely asymptotic. One of the sources of divergent behaviour, 
called IR renormalon \cite{tHooft:1977xjm,Parisi:1978az,David:1983gz,Mueller:1984vh,David:1985xj}, arises from the sensitivity of the process to the 
inevitable long-distance scale $\Lambda_{\rm QCD}$. The degree of 
IR sensitivity limits the ultimate accuracy of the perturbative 
approximation. For the pole mass of a heavy quark, 
this observation has been exceptionally important for particle 
physics phenomenology, leading to a better understanding of quark 
mass renormalization at scales of order and below the mass of 
the quark, and to much improved precision in heavy quark and 
quarkonium physics. 

The quark two-point function 
\begin{equation}
\int d^4 x \,e^{i p x}\,\langle\Omega|T(q_{a}(x)\bar{q}_b(0))
|\Omega \;\rangle\, 
\stackrel{p^2\to m^2}{\rightarrow} 
\delta_{ab} \,Z \, \frac{i (\not \!p +m)}{p^2-m^2} + 
\mbox{less singular} 
\end{equation}
has a pole\footnote{Due to the masslessness of the gluon, there 
is no gap between the single-particle pole and the multiparticle 
cut starting at $p^2=m^2$, and the residue $Z$ is IR divergent.} 
to any order in the perturbative expansion. The location of 
the pole in the complex $p^2$ plane defines the pole mass of the 
quark, $m$. The pole is shifted off the real axis by a small amount 
due to the weak 
decay of the quark, but for the discussion in this article the imaginary 
part is not relevant and hence will be ignored. The pole mass of a quark 
is IR-finite \cite{Tarrach:1980up,Kronfeld:1998di}. It can 
be related to other renormalized quark mass definitions  
order by order in perturbation theory. It is nevertheless not physical, 
as quarks do not exist as free, asymptotic particle states and 
the scattering matrix of QCD does not exhibit a pole at $m^2$. 

It is intuitively obvious that the strong IR physics of QCD, which 
is not captured by perturbation theory, should contribute an 
amount $\Lambda_{\rm QCD}$ to hadron masses. For a meson state 
$M$ composed of $\bar{q}_i q_j$, 
\begin{equation}
\label{mesonmass}
M=m_i+m_j + \mbox{const} \times \Lambda_{\rm QCD} + \ldots,
\end{equation}
which renders the notion of the pole mass useless for light 
quarks with masses $m_i \lesssim \Lambda_{\rm QCD}$. However, for 
mesons containing heavy quarks with $m_Q\gg  \Lambda_{\rm QCD}$, 
the pole mass provides a first approximation to the meson mass up to 
power corrections of relative order $\Lambda_{\rm QCD}/m_Q$. 
Starting from, say, the $\overline{\rm MS}$ mass of the heavy quark, 
the pole mass represents the perturbatively calculable 
leading-power approximation to the meson mass, 
just as the perturbative calculation  in terms of quarks and gluons of 
the total  $e^+ e^- \to $~hadrons cross section at high energy 
does to the physical hadroproduction cross section.

There is a deep connection between power corrections and IR renormalon 
divergence of the QCD perturbative expansion. The existence of a 
{\em linear} power correction is related to the strong 
IR renormalon divergence of the pole mass series that was 
discovered in \cite{Beneke:1994sw,Bigi:1994em} and is the 
subject of this article. In consequence, while the pole mass 
of a quark appeared as the natural choice for processes involving 
nearly on-shell heavy quarks, the concept has since largely been 
abandoned in precision calculations in favour of alternative, (leading) 
renormalon-free mass definitions.

\section{Pole mass series}

\subsection{Basic definitions}

The pole mass $m$ is related to the  \MSB{} mass by
\begin{eqnarray}
\mpole &=& \mr(\mum) \bigg(1+\sum_{n=1}^\infty c_{n}(\mu,\mum,m(\mum)) 
\,\alpha_s^{n}(\mu) \bigg)   \,.
\label{eq:poleMSbar} 
\end{eqnarray}
Here $\as(\mu)$ is the $\overline{\rm MS}$ coupling at scale $\mu$ 
in QCD with $n_l$ light quarks, and  $\mr(\mum)$ stands for the heavy 
quark \MSB{} mass evaluated at the scale $\mum$. 
In the following I will often set $\mu_m=\mB$, where $\mB$ refers to 
the \MSB{} mass, evaluated self-consistently at the scale equal to the 
mass itself, i.e.
\begin{equation}
\mB=\mr(\mB)\,.
\end{equation}

We shall see below that the series \eqref{eq:poleMSbar} diverges 
for any value of $\alpha_s\not=0$. Although a proof does not exist for 
QCD, it is reasonable to assume that it is asymptotic, and 
approximates the heavy-meson mass up to exponentially small terms 
in $\alpha_s$, equivalent to power corrections in $\lambdaQCD/\mB$.  
Asymptotic expansions can sometimes be summed using the Borel transform. 
Given a power series
\begin{equation}
f(\as)=\sum_{n=1}^{\infty} c_n \as^n\,,
\end{equation}
the corresponding Borel transform is defined by
\begin{equation}
  B[f](t)=\sum_{n=0}^{\infty} c_{n+1}\,\frac{t^n}{n!}\,.
\end{equation}
By convention, the tree-level term ``1'' in \eqref{eq:poleMSbar} is 
excluded from the definition. A factorially divergent series of the 
form 
\begin{equation}
r_n = K a^n \Gamma(n+1+b)
\label{asymptotic}
\end{equation}
has the Borel transform
\begin{equation}
\label{borelsingularity}
B[f](t) = \frac{K \Gamma(1+b)}{(1-a t)^{1+b}}
\end{equation}
with a singularity at $t=1/a$. The Borel integral 
\begin{equation}
I[f]\equiv \int_0^\infty dt\,e^{-t/\as}\, B[f](t)
\label{If}
\end{equation}
has the same series expansion as $f(\as)$ and provides the exact result 
under suitable conditions. However, for our case of interest, there 
will be a singularity on the integration contour, rendering the Borel 
integral as given ill-defined. Deformations of the contour around the 
pole or branch cut, or the principal-value prescription, result 
in the ambiguity 
\begin{equation}
\delta f \equiv |\mbox{Im}\,I[f]| = \frac{\pi |K|}{a}\,e^{-1/(a\as)}\,
(a\as)^b\qquad\mbox{(if $a>0$)} 
\label{ambiguity}
\end{equation}
of the 
Borel integral, which has the form of a power correction 
proportional to $\lambdaQCD^{-2\beta_0/a}$. The QCD $\beta$-function 
is defined as  $\beta(\alpha_s) 
\equiv \mu^2\partial\alpha_s(\mu)/\partial\mu^2 = 
\beta_0 \alpha_s^2+\beta_1\alpha_s^3+\ldots$ with 
$\beta_0 = -(11-2 n_f/3)/(4\pi)$. This 
ambiguity provides a quantitative measure of the limit to the 
accuracy of a purely perturbative calculation.

The linear power correction to the pole mass therefore 
corresponds to $a=-2\beta_0$. More generally, the pole mass can be 
regarded as the first term 
of an asymptotic expansion of the meson mass in powers of 
$\alpha_s$ and $\lambdaQCD/\mB$, which in modern language has a 
trans-series structure (again, no proof).\footnote{See 
the reviews \cite{Beneke:1998ui,Beneke:2000kc} for 
the interpretation of IR renormalons in the 
context of short-distance and operator product expansions and 
more formalism.}

\subsection{Linear IR sensitivity and the large-$n_f$ approximation}

To gain intuition, we start from the leading IR renormalon 
divergence of the one-loop correction to the pole mass with fermion-loop 
insertions into the gluon line, often referred to as the large-$n_f$ 
approximation. The relevant expression is 
\begin{eqnarray}
\Delta m \equiv m-\mr(\mu) &=& 
(-i)g_s^2 C_F \mu^{2\epsilon}
\int\!\frac{d^d k}{(2\pi)^d}\,\frac{\gamma^\mu 
(\!\not\!p+\!\not\!k+m)\,\gamma^\nu}{k^2\,((p-k)^2-m^2)}\big|_{p^2=m^2} 
\left(g_{\mu\nu}-\frac{k_\mu k_\nu}{k^2}\right)
\nonumber\\
&&\times\,\sum_{n=0}^\infty 
\left[\beta_0\alpha_s 
\ln\left(\frac{-k^2 e^{-5/3}}{\mu^2}\right)\right]^{\! n} 
+ \mbox{counterterms}
\label{polm}
\end{eqnarray}
The all-order $\MSB$ counterterms can be found in \cite{Beneke:1994sw}. 
They do not diverge factorially, and can therefore be ignored when 
discussing the large-$n$ behaviour. Strictly speaking, 
the fermion-loop insertions provide only the $n_f$-dependent 
part of $\beta_0$ in \eqref{polm}. In full QCD, consistency of 
the trans-series interpretation of short-distance expansions 
requires the full expression for $\beta_0$, as will be seen 
below. The diagrammatic recovery of the full $\beta_0$ is 
discussed in \cite{Beneke:1998ui}. The substitution of the 
full $\beta_0$ in fermion bubble-chain diagrams is often 
referred to as ``naive non-abelianization'' 
\cite{Broadhurst:1994se,Beneke:1994qe}.

For $p^2=m^2$ the integral scales as $d^4 k/k^3$ for small $k$. 
It is thus IR finite, but the contributions from $k$ smaller than 
$\lambdaQCD$, where perturbation theory is not valid, is of 
order $\lambdaQCD$. That this should imply that the pole mass cannot 
be defined to better accuracy than $\mathcal{O}(\lambdaQCD)$ 
was noted in \cite{Bigi:1993zi}. The connection to the IR renormalon 
divergence of the perturbative expansion was established shortly 
after \cite{Beneke:1994sw,Bigi:1994em}. Indeed, the increasing 
power of logarithms in \eqref{polm} enhance the IR region and 
yield (after Wick rotation)
\begin{equation}
\int_0^\lambda dk\,\ln^n\!\left(\frac{k^2}{\mu^2}\right) 
\stackrel{n\gg 1}{=} (-2)^n n!\,
\end{equation}
with typical $k\sim \mu \,e^{-n}$. One can also take the Borel 
transform of \eqref{polm}, sum the series, which yields an 
effective gluon propagator. The exact expression for the 
Borel transform of $m-\mr(\mu)$ can be found in 
\cite{Beneke:1994sw}. Approximating the integrand to its leading term 
in the small-$k$ behaviour is sufficient to obtain the dominant 
IR renormalon singularity at  $t=-1/(2 \beta_0)$ 
(closest to the origin of the Borel plane), 
resulting in 
\begin{equation}
B[\Delta m] = \frac{C_F e^{5/6}}{\pi}\,
\mu\,\frac{1}{1+2\beta_0 t}\,.
\end{equation}
The corresponding 
asymptotic behaviour of the 
series expansion in $\alpha_s=\alpha_s(\mu)$ is 
\begin{equation}
\label{dif}
m_{\rm pole}-m_{\overline{\rm MS}}(\mu) = 
\frac{C_F e^{5/6}}{\pi}\,
\mu\,\sum_n (-2\beta_0)^n\,n!\,\alpha_s^{n+1}.
\end{equation}
As expected, with $a=-2\beta_0>0$ 
the ambiguity of the Borel integral and hence of the 
pole mass\footnote{Implicitly, it is assumed 
that the $\overline{\rm MS}$ mass has no IR sensitivity, 
since it is essentially the bare mass up to pure UV poles.}
 is proportional 
to $\lambdaQCD$, independent of the arbitrary renormalization scale 
$\mu$. 

\subsection{Exact characterization of the divergence}

The relation of IR and ultraviolet (UV) renormalon divergence with 
the small- and large-momentum behaviour of Feynman diagrams, 
respectively, allows for a precise characterization of the 
corresponding singularities in the Borel transform in terms of 
the factorization properties of observables and correlation 
functions in these limits. In asymptotically free, renormalizable 
field theories the UV renormalon singularities occur at 
$t=n/\beta_0<0$ and can be related to local operators of 
dimension $4+2 n$ in the regularized theory expanded for 
large values of a dimensionful cut-off \cite{Parisi:1978bj} (see 
\cite{Beneke:1997qd} for the case of QCD). If the theory has 
power UV divergences, as is usually the case for effective field 
theories, the UV renormalon singularities extend into the 
positive real axis of the Borel plane. Likewise, the IR behaviour 
of correlation functions is often amenable to expansions in the ratio of 
$\lambdaQCD$ and a hard scale, in which case the IR renormalons 
at $t=-n/\beta_0>0$ are related to higher-dimensional terms in 
the operator product expansion (OPE). For example, for the 
two-point function of two vector currents, 
\begin{equation}
\label{eq:ope}
\Pi(Q) = C_0(\alpha_s,Q/\mu) + \frac{1}{Q^4}\,C_{GG}(\alpha_s,Q/\mu)\,
\langle \frac{\alpha_s}{\pi} G^2\rangle(\mu) + O(1/Q^6),
\end{equation}
the position $t=-2/\beta_0$ of the leading IR renormalon divergence 
of its perturbative series $C_0(\alpha_s,Q/\mu)$ is determined 
by the dimension (four) of the gluon condensate correction
\cite{Parisi:1978az,David:1983gz,Mueller:1984vh,David:1985xj}. 
For both, UV and IR renormalons, the parameter $b$ in 
\eqref{borelsingularity} is determined by the anomalous dimension(s) 
of the relevant operators. Through renormalization-group 
equations (RGEs), one can determine the $\alpha_s$ dependence of the 
ambiguity of the Borel integral and thus determine $1/n$ corrections 
to the leading large-order behaviour in terms of OPE coefficient 
functions, the anomalous dimensions of all operators of a given 
dimension, and the beta function coefficients 
\cite{Beneke:1993ee}.\footnote{Explicit formulae can be found in  
\cite{Beneke:1998ui,Beneke:2000kc}.}
This leads to the remarkable conclusion that the singular points 
of the Borel transform due to IR and UV renormalons can be 
completely specified, except for a set of normalization constants 
$K$ in \eqref{borelsingularity}, whose number matches (at most) 
the number of 
operators. They appear as initial conditions of the RGE and should 
be viewed as non-perturbative \cite{Beneke:1993ee,Grunberg:1992hf}.

The application of these ideas to the large-order behaviour of the 
pole mass series exhibits some unique features: a) due to the linear 
IR sensitivity, the IR renormalon divergence is particularly strong 
and dominates over the sign-alternating UV renormalon series; b) 
the leading IR renormalon singularity at $t=-1/(2\beta_0)$ involves 
only a single operator and therefore a single unknown normalization 
constant; c) the operator has vanishing anomalous dimension, hence 
the sub-asymptotic $1/n^k$ corrections are determined only from the 
QCD beta-function, which is known to high-order in perturbation 
theory. 

The derivation of these statements in \cite{Beneke:1994rs} builds 
on the observation \cite{Beneke:1994sw} that the leading IR 
renormalon in the pole mass is related to an UV renormalon pole 
at the same position $t=-1/(2\beta_0)$ of the self-energy 
$\Sigma^{\rm static}$ of the static quark field in 
heavy quark effective theory (HQET) with Lagrangian 
\begin{equation}
\mathcal{L}_{\rm eff} = \bar{h}_v iv\cdot D h_v + 
\mathcal{L}_{\rm light}\,.
\end{equation} 
This UV renormalon pole exists, because in contrast to full 
QCD, the static self-energy is linearly UV divergent. 
The only operator with the required mass dimension three is 
$\bar{h}_v h_v$. It follows that the imaginary part of 
the Borel integral of $\Sigma^{\rm static}$ is given by 
\begin{equation}
\mbox{Im}\,I[\Sigma^{\rm static}](\alpha_s,p,\mu) = 
E(\alpha_s,\mu)\,\Sigma^{\rm static}_{\bar{h} h}(\alpha_s,p,\mu),
\end{equation}
where 
$\Sigma^{\rm static}_{\bar{h} h}$ is the static self-energy with a 
zero-momentum insertion of $\bar{h}_v h_v$. The coefficient $E(\alpha_s,\mu)$ 
satisfies the RGE
\begin{equation}
\left(\mu^2\frac{\partial}{\partial\mu^2}+\beta(\alpha_s)
\frac{\partial}{\partial \alpha_s}-\gamma_{\bar{h}_v h_v}(\alpha_s)
\right) E(\alpha_s,\mu) = 0\,.
\end{equation}
However, $\gamma_{\bar{h}_v h_v}$ vanishes to all orders in 
perturbation theory, since $\bar{h}_v h_v$ is the conserved 
heavy quark number current of HQET. This justifies statements 
b) and c). One then 
shows that \cite{Beneke:1994rs}
\begin{equation}
\mbox{Im}\,I[\Delta m] = -E(\alpha_s,\mu) = 
{\rm const}\times\mu\,\exp\left(\int_{\alpha_s} \!
dx\frac{1}{2\beta(x)}\right) 
= {\rm const}\times \lambdaQCD.
\end{equation}
The $\alpha_s$-dependence of the imaginary part of the Borel 
integral determines the large-order behaviour of the perturbative 
expansion of $\Delta m$ according to (\ref{asymptotic}), (\ref{If}) 
up to a single normalization constant, $N$.
Defining
\begin{eqnarray}
\label{eq:cnasymp}
c_{n}(\mu,\mu_m,\mr(\mum)) & \underset{n\to\infty}\longrightarrow & 
N \ac_{n}(\mu,\mr(\mum)) \equiv 
N \frac{\mu}{\mr(\mum)}\,\acb_{n}\,, 
\end{eqnarray}
the result is 
\begin{eqnarray}
\label{poleasymp}
\tilde{c}_{n+1}^{(\rm as)} &=& (-2\beta_0)^n\,
\frac{\Gamma(n+1+b)}{\Gamma(1+b)} 
\Bigg[1+\frac{s_1}{n+b}+\frac{s_2}{(n+b)\,(n+b-1)}
\nonumber\\
&&+\,\frac{s_3}{(n+b)\,(n+b-1)\,(n+b-2)} + \ldots\Bigg],
\end{eqnarray}
where 
\cite{Beneke:1994rs,Beneke:2016cbu}
$b=-\beta_1/(2\beta_0^2)$ and 
\begin{eqnarray}
\label{eq:ctildenasympparams}
s_1 &=& \left(-\frac{1}{2\beta_0}\right)\left(-\frac{\beta_1^2}{2
\beta_0^3}+\frac{\beta_2}{2\beta_0^2}\right),
\\[0.2cm]
s_2 &=& \left(-\frac{1}{2\beta_0}\right)^2\left(\frac{\beta_1^4}{8
\beta_0^6}+\frac{\beta_1^3}{4\beta_0^4}-\frac{\beta_1^2\beta_2}{4
\beta_0^5}-\frac{\beta_1\beta_2}{2\beta_0^3}+\frac{\beta_2^2}{8
\beta_0^4}+\frac{\beta_3}{4\beta_0^2}\right),
\\[0.2cm]
s_3 &=& 
\left(-\frac{1}{2\beta_0}\right)^3\bigg(
-\frac{\beta_1^6}{48 \beta_0^9}
-\frac{\beta_1^5}{8\beta_0^7}
-\frac{\beta_1^4}{6 \beta_0^5}
+\frac{\beta_1^4\beta_2}{16 \beta_0^8}
+\frac{3 \beta_1^3 \beta_2}{8 \beta_0^6}
+\frac{\beta_1^2 \beta_2}{2 \beta_0^4}
-\frac{\beta_1^2 \beta_2^2}{16 \beta_0^7}
\nonumber\\
&& 
-\frac{\beta_1^2 \beta_3}{8 \beta_0^5}
-\frac{\beta_1 \beta_2^2}{4 \beta_0^5}
-\frac{\beta_1 \beta_3}{3 \beta_0^3}
+\frac{\beta_2^3}{48\beta_0^6}
-\frac{\beta_2^2}{6\beta_0^3}
+\frac{\beta_2 \beta_3}{8 \beta_0^4}
+\frac{\beta_4}{6 \beta_0^2}
\bigg)\,.
\end{eqnarray}
The pole mass series is particularly simple, because the 
large-order behaviour is completely determined 
in terms of the $\beta$-function coefficients. Since the 
five-loop beta-function coefficient $\beta_4$ is now known 
\cite{Baikov:2016tgj,Herzog:2017ohr,Luthe:2017ttg}, the 
sub-asymptotic behaviour including $1/n^3$ corrections 
is known. Numerically, for the interesting cases discussed 
below, the corrections to the leading large-$n$ behaviour 
do not exceed 3\% for $c_4$.

\subsection{The top, bottom and charm mass series}

The normalization constant $N$ in (\ref{eq:cnasymp}) cannot 
be determined exactly by purely perturbative methods. However, 
given that the pole-\MSB~mass relation \eqref{eq:poleMSbar} is 
known to the four-loop order \cite{Marquard:2015qpa,Marquard:2016dcn} 
and the asymptotic behaviour is known including $1/n^3$ 
corrections, one may attempt to match the two at $n=4$. 
In other words, while 
\begin{eqnarray}
\label{eq:Ndet}
N&=&\lim_{n\to\infty} \frac{c_{n}(\mu,\mum,\mr(\mum))}
{\ac_n(\mu,\mr(\mum))}\,,
\end{eqnarray}
we evaluate the ratio for $n=4$ and check the stability of 
the result by comparison with $n=3$ \cite{Beneke:2016cbu}. 
In the following, all $\nl$ quarks other than the heavy 
quark will be assumed to be massless. The effect of internal 
quark mass effects will be discussed below. The coupling 
$\alpha_s$ in \eqref{eq:poleMSbar} is the \MSB~coupling in the 
$n_l$-flavour theory. Since the IR theories are different, 
$N$ is expected to depend on $\nl$.

It is interesting to apply this method to the large-$\nl$ limit 
(more precisely, $\nl\to -\infty$). In this limit, $b=s_i=0$, 
and $N$ can be calculated exactly from \eqref{dif} to be
\begin{equation}\label{eq:Nlargenl}
\lim_{|\nl| \to \infty} N=\frac{\cf}{\pi} \times e^{\frac{5}{6}}\,,
\end{equation}
which equals  $0.97656$ ($\cf=4/3$ for  $N_c=3$). This can be 
compared to evaluating \eqref{eq:Ndet} for $n=4$ at $\mu=\mu_m=
\mB$, which gives $0.971$ in very good agreement with the exact 
result. According to \eqref{eq:cnasymp} the dependence of the asymptotic 
behaviour on $\mu$ and $\mu_m$ is very simple, since the logarithms of 
$\mu$ in perturbation theory must exponentiate to powers 
asymptotically. The approximate determination of $N$ is most accurate 
when choosing $\mu\approx\mB$ and exhibits a plateau around this 
value \cite{Beneke:2016cbu}.

With this validation of the method we consider the pole to 
\MSB~mass series 
for the top, bottom and charm quark, corresponding to $\nl=5,4,3$, 
respectively. For a detailed analysis of the top-quark case, 
see~\cite{Beneke:2016cbu}. Setting  $\mu=\mu_m=
\mB_Q$ for $Q=t,b,c$, one finds
\begin{equation}
\label{norm}
N = 0.4606 \mbox{ (top)},~0.5048 \mbox{ (bottom)},~0.5366 \mbox{ (charm)}\,,
\end{equation}
where the tiny difference relative to \cite{Beneke:2016cbu} is due 
to the inclusion of $s_3$, which was 
not fully available at the time (lack of $\beta_4$). A conservative 
estimate of the uncertainty of these values from the independent 
variations of $\mu$ and $\mu_m$ is $\pm 10\%$ (error symmetrized, 
see \cite{Beneke:2016cbu}), but the accuracy of the large-$n_l$ 
result at  $\mu=\mu_m=\mB$ suggests that it may be considerably 
smaller. It is worth noting that $N$ is only half as large 
as the large-$\nl$ result for physical values of $\nl$, 
implying that the intrinsic ambiguity of the pole mass is 
smaller than inferred from the one-loop correction dressed by 
fermion loops. 

\begin{figure}
\centering
\includegraphics[width=0.49\textwidth]{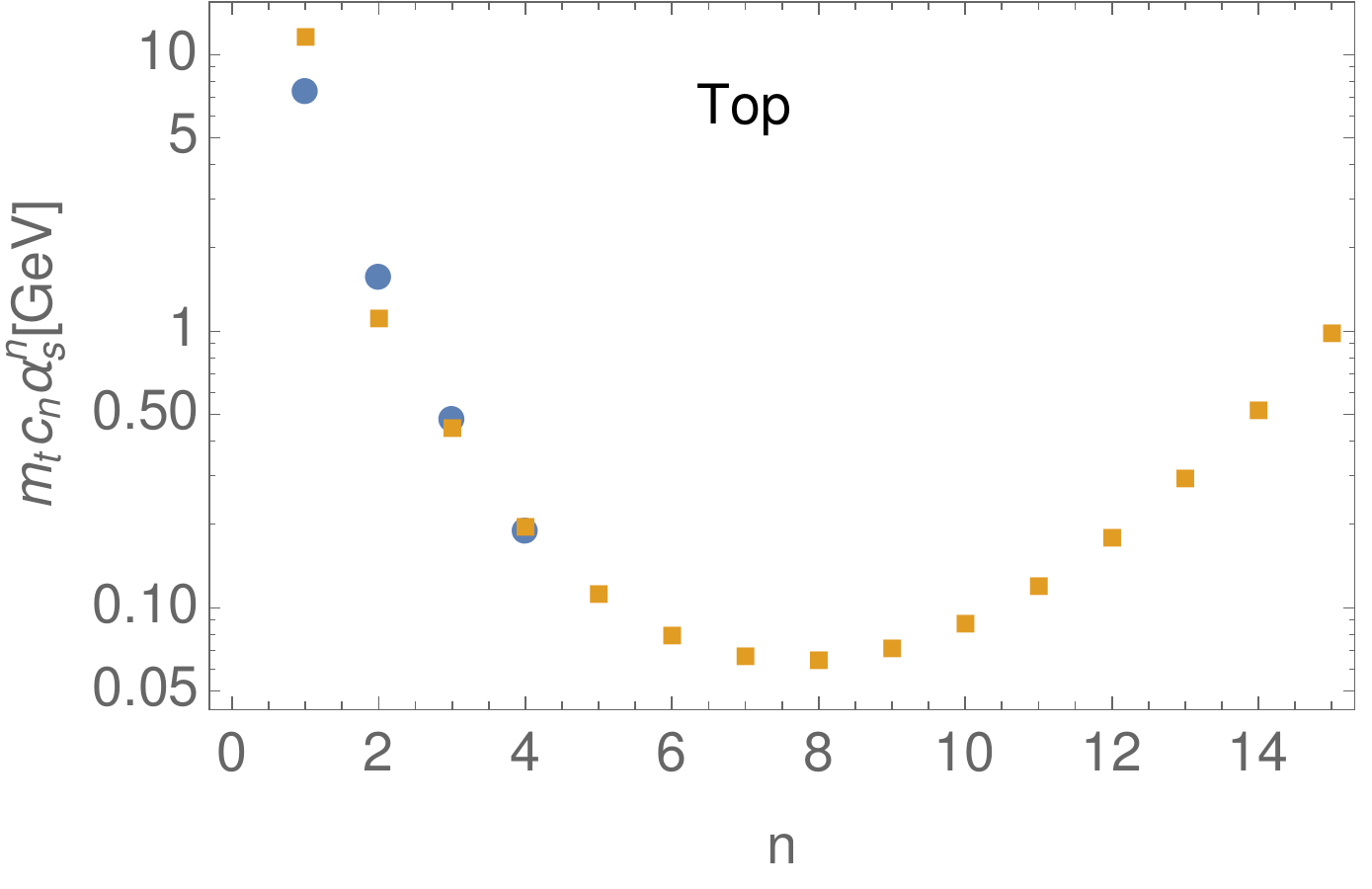} 
\includegraphics[width=0.49\textwidth]{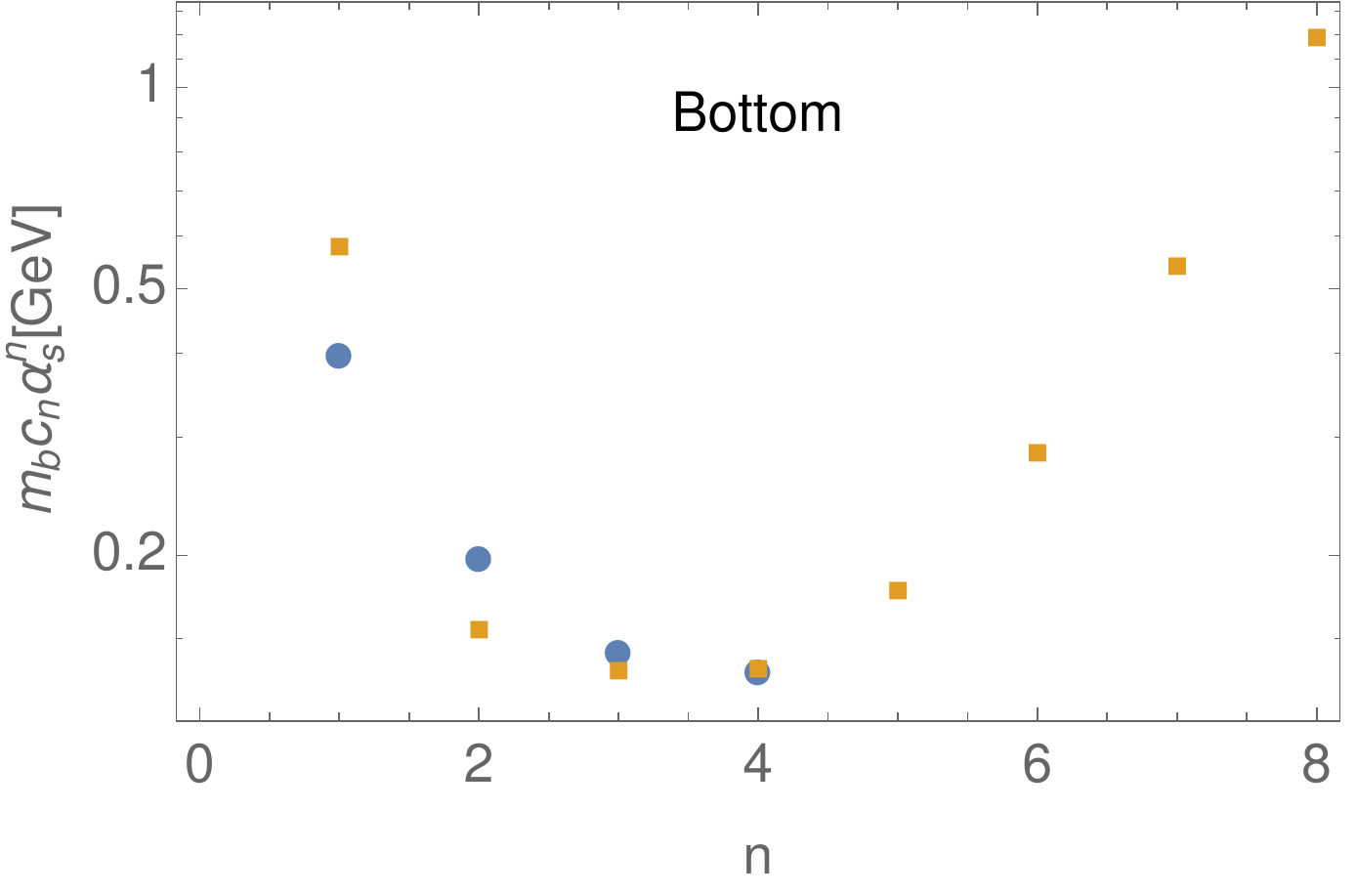} 
\vskip0.2cm
\includegraphics[width=0.49\textwidth]{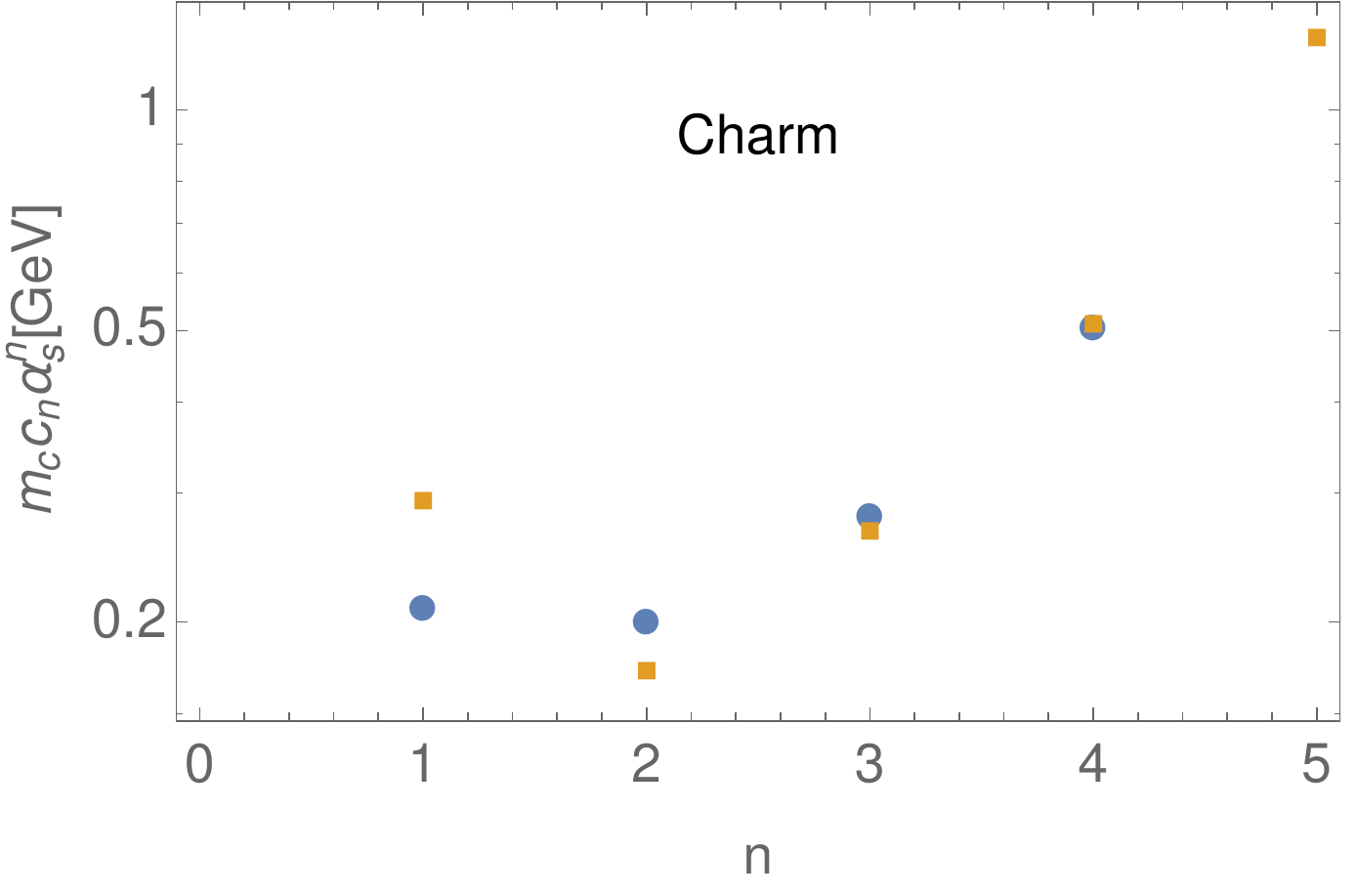} 
\caption{Graphical representation of the series \eqref{mtseries} 
-- \eqref{mcseries}, exhibiting the typical dependence of the 
size of terms with order of a factorially divergent series. The 
(blue) circles for $n<5$ represent exactly known terms, the 
(yellow) squares the asymptotic formula applied to all $n>0$.}
\label{fig:series}      
\end{figure}

To display the numerical properties of the series, I 
use the \MSB~mass values $\mB_t=163.643\,$GeV, $\mB_b=4.20\,$GeV and 
$\mB_c=1.28\,$GeV. The strong coupling is taken to be 
$\alpha_s^{(5)}(m_Z)=0.1180$ at the scale $m_Z=91.1876~$GeV and 
evolved with five-loop accuracy to $\mB_Q$ 
including the flavour thresholds at $2\mu_b = 9.6\,$GeV and 
$2\mu_c=3\,$GeV with the help of {\tt RunDec} 
\cite{Chetyrkin:2000yt,Herren:2017osy}. The series coefficients 
are then evaluated at $\mu=\mu_m=\mB_Q$ in an expansion 
in $\alpha_s^{(5)}(\mB_t)=0.1084$, $\alpha_s^{(4)}(\mB_b)=0.2246$, 
$\alpha_s^{(3)}(\mB_c)=0.3889$, for top, bottom and charm, 
respectively. The result is
\begin{eqnarray}
\label{mtseries}
m_t &=& 163.643+7.531+1.606+0.494 + 0.194 + {\it 0.111 + 0.079} 
\nonumber\\
&&{\it +0.066+{\bf 0.064}+0.070+0.087+0.119+0.178}
+\ldots\,{\rm GeV}\\[0.3cm]
\label{mbseries}
m_b &=& 4.200+0.400+0.199+0.145 + {\bf 0.135} + {\it 0.177 + 0.284}
\nonumber\\ 
&&{\it +0.539+1.185}+\ldots\,{\rm GeV}\\[0.3cm]
\label{mcseries}
m_c &=& 1.280+0.211+{\bf 0.202}+0.282 + 0.510 + 
{\it 1.259 + 3.798}+\ldots\,{\rm GeV}
\end{eqnarray}
for the series expansion of the mass conversion formula. In these 
expressions the first five numbers correspond to the five exactly 
known terms including the four-loop order, and the subsequent 
numbers in italics are obtained from the asymptotic formula 
\eqref{poleasymp}. By construction, the asymptotic formula agrees 
with the exact one for the fifth term. The minimal term 
of the series is highlighted in bold face. The asymptotic formula 
corresponds to the ``prediction''
\begin{equation}
c_5 =  45.43 \mbox{ (top, $\nl=5$)},~73.69 \mbox{ (bottom, $\nl=4$)},~110.56 \mbox{ (charm, $\nl=3$)}
\end{equation}
at $\mu=\mu_m=\mB_Q$ for the presently unknown five-loop conversion 
coefficient. The behaviour of the series is illustrated in 
Figure~\ref{fig:series}, including an extrapolation of the 
asymptotic formula to all $n>0$. It is apparent that it is already 
accurate at the three-loop order, $n=3$.

Comparing \eqref{mtseries} -- \eqref{mcseries}, we observe that 
the top mass series attains its smallest term at the eighth order 
in perturbation theory, far beyond the four-loop order currently 
known. On the other hand, the bottom series reaches its minimal 
term at this order, while the charm series starts to diverge from 
the two-loop order, which renders the charm pole mass of limited use 
for phenomenology. From a pragmatic point of view, the minimal 
term represents the ultimate accuracy beyond which the purely 
perturbative use of the pole quark mass ceases to be meaningful. 
The minimal term scales as $\sqrt{\alpha_s(\mB_Q)} \,\lambdaQCD$ and 
decreases with larger $m_Q$, which reflects 
the fact that the minimum is shallower in this case. A 
renormalization-group invariant measure of the intrinsic limitations 
of the concept of the pole mass can be defined in terms of the 
ambiguity \eqref{ambiguity} of the Borel integral of the series, 
\begin{equation}
\label{mQambiguity}
\delta m_Q = \frac{\pi N}{|2\beta_0|}\times\lambdaQCD^{(n_f)}\,,
\end{equation}
exactly proportional to $\lambdaQCD$, as it should be. Dividing 
by $\pi$ gives a numerical value close to the minimal term for the 
top mass series, and this definition of ultimate accuracy has been 
adopted in \cite{Beneke:2016cbu}.

Determining the non-perturbative normalization $N$ of the 
leading pole mass renormalon singularity from matching to the highest known 
order is perhaps the simplest and most intuitive, but not the 
only method that has been suggested. I refer to 
\cite{Pineda:2001zq,Lee:2002sn,Lee:2003hh,Hoang:2008yj,Ayala:2014yxa,Lee:2015owa,Hoang:2017suc,Sumino:2020mxk} for other work, noting that 
earlier work did not have access to four-loop accuracy. 
As will be seen in Sec.~\ref{sec:PS} below, $N$ is related to 
a similar leading renormalon constant of the static heavy quark 
potential by a factor $-1/2$
\cite{Beneke:1998rk,Hoang:1998nz,PinedaPhD}. Some of the quoted 
works apply this relation to infer $N$ from an analysis of 
the series expansion of the colour-singlet Coulomb potential. 
The normalization constants are much more difficult to obtain 
when there is more than one constant, or an interference of 
sign-alternating UV and fixed-sign IR renormalon behaviour, as 
is the case for generic observables. In this case one can resort 
to simplified parameterizations of the Borel transform as was done 
for $\tau$-decay spectral moments \cite{Beneke:2008ad}, but the 
level of rigour and precision that has been achieved for the 
pole mass of a heavy quark is unmatched by any other series 
in QCD.

\subsection{Internal quark mass effects}

The analysis assumed up to now that the lighter quarks are massless. 
In low orders of perturbation theory, this is often a good approximation, 
especially for the top quark pole mass, where $m_c, m_b\ll 
m_t$. (We do not consider the effect of the up, down and strange 
quark mass and always neglect quark masses smaller than $\lambdaQCD$.) 
However, in the regime where the series is dominated by the leading
renormalon divergence, the typical loop momentum at order 
$\alpha_s^{n+1}$ is of order $m_Q e^{-n}$. Internal quark mass effects 
from the bottom and charm quark are expected to become important in higher 
orders. The minimal term of the series is attained when the typical 
loop momentum is of order $\LambdaQCD$. At this scale the theory 
is a theory of three massless flavours, independent of whether 
the heavy quark was the top, bottom or charm quark. Hence, the true 
large-$n$ behaviour of the series beyond the minimal terms is 
always determined by the $\nl=3$ result, and likewise the ambiguity 
(\ref{mQambiguity}) should involve the
$\Lambda$-parameter $\LambdaQCD^{(3)}$ in the three-flavour scheme, 
always excluding the bottom and charm quark, independent of~$\nl$. 

The decoupling of internal quarks with masses $m_q$ larger than 
$\lambdaQCD$ in the renormalon asymptotic behaviour was studied 
analytically and numerically in the large-$n_l$ limit 
\cite{Ball:1995ni}, and the described behaviour has been 
demonstrated. More precisely, the analysis showed that the asymptotic 
behaviour of the series in a 
theory with $n_l$ quarks of which $n_m$ are massive, approaches the series 
of the theory with $n_l-n_m$ massless quarks when both are expressed in 
terms of the $\overline{\rm MS}$ coupling $\alpha_s^{(n_l-n_m)}(m_Q)$ in 
the $n_l-n_m$ flavour scheme. However, as noted above the 
large-$n_f$ limit overestimates the normalization of the leading 
renormalon by about a factor of two. Furthermore, one is rarely interested 
in the formal large-$n$ behaviour of the series beyond the 
minimal term, but rather in 
the approach to it. At such intermediate orders, the typical 
loop momentum crosses the flavour thresholds, as the order of 
perturbation theory increases, and the internal quark masses 
are neither negligible nor decoupled. 

The issue is especially relevant for the top quark, since the masses of 
the bottom and charm quark are too small in relation to $m_t$ to express 
the entire series in terms of the four- or three-flavour coupling. 
In contrast, one may argue \cite{Ayala:2014yxa} that the bottom pole 
to \MSB~mass conversion factor should be expressed in terms of 
$\alpha_s^{(3)}(m_b)$ rather than the four-flavour coupling 
$\alpha_s^{(4)}(m_b)$. For the two- and three-loop 
coefficients, for which the mass dependence is known 
\cite{Gray:1990yh,Bekavac:2007tk,Fael:2020bgs}, this substitution 
indeed renders the charm mass effect almost negligible. 
A quantitative investigation of bottom and charm mass effects 
on the top pole mass series was undertaken in 
\cite{Beneke:2016cbu,Hoang:2017btd}. The following discussion 
is adapted from \cite{Beneke:2016cbu}.

I first recall that the numerical series 
\eqref{mtseries} -- \eqref{mcseries} for $m_Q$ include internal 
loops of $Q$, but are expressed in terms of 
$\alpha_s^{(\nl)}(\mB_Q)$, where $\nl$ is the number of 
massless quarks, including bottom and charm for the case of 
$Q=\,$top ($\nl=5$). To estimate the effect of the finite 
bottom and charm mass, we switch from the five- to the four-flavour 
scheme at the order, where the typical internal loop momentum is of 
order $m_b$, which is ${\cal O}(\alpha_s^5)$, and from the four- to 
the three-flavour scheme at ${\cal O}(\alpha_s^6)$. Since the mass 
effect is not known for $c_4$ at the four-loop order, and since 
$c_n$ beyond the four-loop order can only be estimated assuming 
dominance of the first renormalon (as done above), this implies 
the following procedure: a) at two and three loops, we include the 
known mass dependence, but $c_4$ is approximated by the massless value. 
For given top $\overline{\rm MS}$ mass, this increases the top pole mass by 
11 (2-loop) + 16 (3-loop) MeV, adopting $\mB_b=4.2$~GeV 
and $\mB_c=1.28$~GeV, out of which 8.1 (2-loop) + 11.2 (3-loop) MeV are 
due to the finite bottom mass, and 2.4 (2-loop) + 4.6 (3-loop) from 
charm.\footnote{The poor convergence of the series of mass corrections 
is expected, since the leading internal mass correction is linear. 
This linear dependence is a consequence of the same linear 
IR sensitivity that causes the leading IR renormalon divergence. 
It is therefore possible to calculate the linear internal mass 
correction by focusing on the soft region of the loop integral, 
which would provide a more accurate estimate of the internal 
mass effect in large orders than the order-dependent decoupling 
procedure employed here.} Since the $c_n$ increase as $n_l$ 
decreases, the mass effect is also expected to be positive in higher orders. 
Hence approximating $c_4$ by its massless value {\em underestimates} the 
mass effect. b) At the five-loop order, we use 
$c_5^{(\rm as)} [\alpha_s^{(4)}(\mB_t)]^5$ with $c_5^{(\rm as)}$ 
determined by matching to the exactly known four-loop coefficient 
for $\nl=4$, that is with normalization $N_m=0.5048$ and beta-function 
coefficients for the four-flavour theory. c) Beyond five loops, 
the remainder of the series is computed with the three-flavour scheme 
coupling $\alpha_s^{(3)}(\mB_t)$ and normalization $N_m=0.5366$. Since
 the bottom and charm quarks are not yet completely decoupled at the five- to 
seven-loop order, and since an extra quark flavour decreases the $c_n$, 
we expect that b) and c) {\em overestimate} the mass effect, since 
the approximation assumes that bottom and charm are already decoupled 
completely. The sum of b) and c) adds another 54 MeV to the top pole 
mass, such that the total mass effect is estimated to be 80~MeV. 
Explicitly, the series \eqref{mtseries} is modified to
\begin{eqnarray}
\label{mtmassiveseries}
m_t &=& 163.643+7.531+1.616+0.510 + 0.194 + {\it 0.140 + 0.106} 
\nonumber\\
&&+\,{\bf 0.094} + {\it 0.096 +0.112+0.145+0.209+0.329}
+\ldots\,{\rm GeV},
\end{eqnarray}
where the increasing importance of finite-quark mass effects with 
order is evident. In case of the top quark pole mass, the 
decoupling of the bottom and charm quark in internal loops 
increases the intrinsic uncertainty of the pole mass concept 
by almost 50\% due to the more rapid divergence of the series in the 
three-massless flavour theory. Note that this ambiguity is independent of 
the precise value of the bottom and charm mass, as long as 
$m_b, m_c \gg \lambdaQCD$. This also implies that it is the same 
for any heavy quark, including the bottom quark, since it depends 
only on the infrared properties of the theory, which is QCD with three 
approximately massless flavours.

Since the bottom quark is neither heavy enough to be decoupled in low 
orders, nor light enough to be ignored, where in both cases a massless 
approximation 
can be justified, there is an inherent uncertainty in the above estimate. 
However, as argued above, the errors in the approximations are expected 
to go in opposite directions, hence we consider $(80\pm 30)$~MeV a 
conservative estimate of the total internal bottom and charm quark mass 
effect on the top pole mass. The 30~MeV error estimate arises from an 
estimate of the neglected mass effect on $c_4$ by extrapolation from the 
known lower orders. The approximation described here has been checked to
work well in models for the series inspired by the large-$n_l$ limit. 

\subsection{Finite width}

With the electroweak interaction turned on, the heavy quarks become 
unstable. In perturbation theory, the pole of the heavy-quark 
propagator is shifted to 
\begin{equation}
m_*^2 = m^2-i m\Gamma(m)\,,
\end{equation}
which defines the pole mass $m$ and on-shell decay width 
$\Gamma$. Unlike the quark mass, the width is not a parameter 
of the Standard Model (SM) --  for heavy quarks it can be computed 
in perturbation theory in terms of $m$ and other SM parameters. 

The width is negligibly small compared to $m$ except for the 
top quark, where $\Gamma_t \approx 1.4~$GeV. The large width 
$\Gamma_t \gg \lambdaQCD$ does not eliminate the renormalon 
divergence of the top pole mass, as was emphasized 
in \cite{Smith:1996xz}. This does not mean that the large 
width is not relevant, since it {\em does} provide a cut-off 
on IR effects for physical observables.  For example, measurements 
on jets containing top quarks are generically linearly 
sensitive to $\lambdaQCD$ and accordingly display a strong 
renormalon divergence, which can be screened by the sizeable 
width \cite{FerrarioRavasio:2018ubr}. In effect, as is intuitive, 
there is simply no quantity, for which the pole mass of a quark 
is ever the relevant parameter, once the quark's width is 
larger than $\lambdaQCD$.

Interestingly, the on-shell width of a quark, $\Gamma(m)$, 
is itself an observable, which is less sensitive to IR physics 
than the pole mass. When the final state 
masses can be neglected, $\Gamma(m) \propto G_F^2 m^5$, where 
$G_F$ denotes the Fermi constant. The leading power corrections 
are of relative order $(\lambdaQCD/m)^2$. However, when the series is 
expressed in terms of the pole mass, an IR renormalon 
divergence indicating an ambiguity of linear order $\lambdaQCD/m$
appears in the series of QCD corrections to the 
tree-level width. This ambiguity is spurious and a consequence 
of using a parameter with stronger IR sensitivity than the 
observable itself. Once the width of the quark is expressed 
in terms of the \MSB~mass or (better) another leading renormalon-free 
mass definition such as will be discussed in Sec.~\ref{sec:newmasses}, 
the leading renormalon is cancelled \cite{Beneke:1994bc}, and the 
series of loop corrections shows a much better behaviour. This 
is particularly important for the decay width of the bottom 
and charm quark, for which $\Gamma$ could otherwise be obtained 
only with large uncertainty. 

\subsection{Beyond the leading renormalon}

Much less is known about the renormalon singularities of the pole 
mass series beyond 
$t=-1/(2\beta_0)$. On general grounds one expects a sign-alternating 
UV renormalon divergence from a singularity at $t=1/\beta_0$, 
and an IR renormalon singularity at $t=-1/\beta_0$ related to the 
$\lambdaQCD^2/m$ kinetic-energy correction to the meson 
mass \eqref{mesonmass}. The Borel transform of the series is 
known exactly in the large-$n_f$ limit \cite{Beneke:1994sw}. Table~11 
in \cite{Beneke:1998ui} displays the breakdown of the 
$n$th order term into the contributions from the first three 
IR renormalon and the first UV renormalon poles, and the 
\MSB~subtraction terms. At least in the large-$n_f$ approximation, 
the subleading poles never contribute more than one permille 
of the dominant asymptotics from $t=-1/(2\beta_0)$ for $n$ 
beyond the four-loop order. For practical purposes, dealing with 
the leading singularity appears to be enough.

Curiously, the Borel transform in the large-$n_f$ limit 
does not exhibit the expected next IR renormalon singularity 
at $t=-1/\beta_0$. The authors of \cite{Beneke:1994sw} speculated 
that Lorentz invariance might forbid a quadratic power-divergent 
mixing of the kinetic energy operator $\bar{h}_v (iD_\perp)^2 h_v$ 
into $\bar{h}_v h_v$, in which case there would be no matrix 
element to compensate the ambiguity of the Borel transform from 
$t=-1/\beta_0$, and hence it should be absent. 
Invoking the virial theorem of HQET, the power-divergent mixing of the 
kinetic energy operator was related to the one of 
$\bar{h}_v i g_s G^{\mu\nu} h_v$ into  
$\bar{h}_v h_v$ \cite{Neubert:1996zy}. This work confirms that 
Lorentz invariance forbids one-loop mixing, which explains the 
absence of the $t=-1/\beta_0$ singularity in the large-$n_f$ 
limit, but also showed that there is no reason for this to 
hold beyond this limit. Recent investigations  \cite{Ayala:2019hkn} of a 
remainder series with the leading renormalon subtracted show 
sign alternation more fitting to UV renormalon behaviour. The 
question whether the subleading renormalon singularity at 
$t=-1/\beta_0$ is absent or simply suppressed by a loop factor, 
is therefore still undecided. 

\section{Renormalon-free ``on-shell'' masses}
\label{sec:newmasses}

Many important properties of heavy quarks are less IR sensitive 
than the pole quark mass itself -- for example, the inclusive decay width 
discussed above, or the production cross section of a heavy 
quark-antiquark pair. Their perturbative expansions do not display 
an IR renormalon singularity at $t=-1/(2\beta_0)$, leading to 
rapid divergence, provided they are not expressed in terms of 
the pole mass. In other words, although the pole mass is IR finite, 
it is for many purposes not a useful renormalized mass parameter. 
Instead, one should use a renormalization convention that is not 
only IR finite but also insensitive to the IR at least at linear 
order $\lambdaQCD$. 

The \MSB~definition suggests itself. 
However, in physical systems where heavy quarks are nearly on-shell 
and have primarily soft fluctuations, the \MSB~mass is not 
the appropriate choice. Being essentially a bare object, 
it does not include the short-distance fluctuations, which 
should have been integrated out to describe soft heavy-quark 
systems. In practice, this means that the \MSB~mass value 
is too far away (by $\mathcal{O}(\mB_Q\alpha_s)$) from the 
pole of the heavy-quark propagator. While the spurious pole 
mass renormalon is eliminated and the asymptotic behaviour improved, 
there still appear large corrections in low orders. It does 
not help to evolve the $\MSB$ mass $\mr(\mu)$ to scales $\mu < \mB$, since 
the $\MSB$ quark-mass anomalous dimension applies only to 
the logarithmic evolution above $\mB$. 

The resolution to the problem consists in quark mass concepts 
that are numerically closer to the pole mass, 
yet are constructed such that their perturbative relation to 
the $\MSB$ mass is free from the leading IR renormalon. This 
has several benefits: 1) The concept is unambiguous, at least 
up to $\mathcal{O}(\alpha_s\lambdaQCD^2/m)$, which is sufficient 
for practical purposes. 2) Such masses can be determined 
accurately from measurements or lattice calculations, and 
3) they can be precisely related to the $\MSB$ mass. 4) The 
impact of light internal quark mass effects is reduced, since 
the leading IR loop momentum contributions have been removed. The 
$\MSB$ mass is then the convenient reference parameter (similar 
to $\alpha_s(m_Z)$ for the strong coupling) to which  
different leading renormalon-free, ``on-shell'' mass definitions can be 
related. 

\subsection{General considerations}

We start from the observation that the asymptotic behaviour 
\eqref{eq:cnasymp} of the pole to $\MSB$ mass conversion  
\eqref{eq:poleMSbar} has a very simple, exact linear dependence 
on the coupling renormalization scale $\mu$, which follows on very general 
grounds \cite{Beneke:1993ee}, as well as on $\mu_m$, which 
appears only through $\mr(\mu_m)$. The asymptotic coefficients 
$\tilde{c}_{n+1}^{(\rm as)}$ themselves in \eqref{poleasymp} are 
$\mr(\mu_m)$, $\mu$ and $\mu_m$ independent. Since the Borel-integral 
ambiguity of the asymptotic series is always $\lambdaQCD$,\footnote{
$\mr(\mu_m)$ is cancelled in $\mr(\mu_m)\,\ac_{n}(\mu,\mr(\mum))$.} we 
can replace $\mr(\mu_m)$ by another scale $\mu_f$. We therefore 
define 
\begin{eqnarray}
 \delta m_X(\mu_f) = \mu_f \sum_{n=1}^\infty s_{n}^X(\mu/\mu_f) 
\,\alpha_s^{n}(\mu) =  \mu_f \sum_{n=1}^\infty s_{n}^X\,\alpha_s^{n}(\mu_f)\,.
\label{delmX}
\end{eqnarray}
The series coefficients $s_n^X(\mu/\mu_f)$ are polynomials 
of order $n-1$ in $\ln(\mu/\mu_f)$ and must be chosen to satisfy
\begin{eqnarray}
\label{eq:snasymp}
s^X_{n}(\mu/\mu_f) & \underset{n\to\infty}\longrightarrow & 
N \frac{\mu}{\mu_f}\,\acb_{n}\,, 
\end{eqnarray}
where $N$ and $\acb_{n}$ are exactly the same as for the coefficients 
in the pole to $\MSB$ mass relation \eqref{eq:cnasymp} and 
\eqref{poleasymp},
respectively.\footnote{It is also assumed that the series 
defines an RGE invariant, such that all logarithms of $\ln(\mu/\mu_f)$ 
can be absorbed into the running coupling at scale $\mu_f$. 
$s_n^X\equiv s_n^X(1)$.}  
Once such $s_n^X$ have been found, we 
can define a leading renormalon-free, ``short-distance'' mass 
$m_X(\mu_f)$ by subtracting $\delta m_X(\mu_f)$ from the pole 
mass $m$:
\begin{eqnarray}
m_X(\mu_f) &\equiv& \mpole - \delta m_X(\mu_f)
\nonumber\\
&=&\mr(\mum) + \sum_{n=1}^\infty 
\left[ \mr(\mum) \,c_{n}(\mu,\mum,m(\mum)) - 
\mu_f \,s_{n}^X(\mu/\mu_f)\right]
\alpha_s^{n}(\mu)  \,.\qquad
\label{poleMSbarsubtracted} 
\end{eqnarray}
By construction the leading IR renormalon divergence of the 
series cancels in the square bracket. 
This in turn guarantees that the series that relates $m_X(\mu_f)$ 
to the $\MSB$ mass $\mr(\mu_m)$ is well-behaved (no leading IR 
renormalon divergence). 

The new scale $\mu_f$ should be chosen such that 
$\lambdaQCD\ll \mu_f\ll m$. The first inequality is required 
for perturbativity. The second guarantees that the 
difference between the pole mass and $m_X(\mu_f)$ is 
only of order $\mu_f\,\alpha_s$ and can be made sufficiently 
small to avoid the problem with the $\MSB$ mass (where 
$\mu_f \sim m$) discussed above. A common feature of 
all renormalon-free quark mass definitions suitable for the 
description of nearly on-shell heavy-quark physics is 
therefore the existence of a new ``subtraction scale'' $\mu_f$ 
and a linear dependence on this scale. This reflects 
that the running of the quark mass changes from logarithmic 
to linear below the scale $m$ in accordance with the fact 
that the self-energy of a point charge is linearly divergent 
in the static or non-relativistic regime and turns logarithmic 
only when the anti-particle fluctuations become relevant in the 
relativistic theory. 

The leading renormalon-free masses satisfy a simple renormalization 
group equation in the subtraction scale $\mu_f$, which may be 
used to relate $m_X(\mu_f)$ at different scales $\mu_{f2}$, 
$\mu_{f1}$, when logarithms of $\mu_{f2}/{\mu_{f1}}$ might have 
to be summed. Defining the anomalous dimension $\gamma_X(\alpha_s)$ 
through
\begin{equation}
\gamma_X(\alpha_s) = - \frac{d m_X(\mu_f)}{d\mu_f}\,,
\end{equation}
the general form \eqref{delmX} of the subtraction term yields 
\begin{eqnarray}
\gamma_X(\alpha_s(\mu_f)) 
&=& \sum_{n=1}^\infty 
s_{n}^X\,\left[\alpha_s^{n}(\mu_f) + 2 n \alpha_s^{n-1}(\mu_f)\,
\beta(\alpha_s(\mu_f))\right]
\nonumber\\
&=& s_1^X \alpha_s(\mu_f) +[s_2^X+2 s_1^X\beta_0]\,
\alpha_s^{2}(\mu_f)+\ldots\,.
\end{eqnarray}

In the following, I discuss several suitable 
mass definitions with no claim to completeness. With respect to 
\eqref{mesonmass}, note that any definition of $\delta m_X(\mu_f)$ 
automatically yields an unambiguous, renormalon-free definition 
$\bar \Lambda_X(\mu_f)$ of the $\bar\Lambda$ parameter that 
appears in the heavy-quark expansion of the heavy meson mass 
and many other HQET expressions by the rearrangement
\begin{equation}
\label{mesonmassmod}
M_Q=[\underbrace{m_Q-\delta m_X(\mu_f)}_{m_X(\mu_f)}]+
[\underbrace{\bar\Lambda+\delta m_X(\mu_f)}_{\bar \Lambda_X(\mu_f)}] 
+ \ldots\,.
\end{equation}
This can be turned around: any renormalon-free 
definition of the HQET parameter $\bar \Lambda$ can be turned into 
a renormalon-free, ``short-distance'', on-shell mass definition.

\subsection{RS mass}

The renormalon subtracted (RS) mass definition \cite{Pineda:2001zq} 
is the first of two schemes, which implement the condition 
\eqref{eq:snasymp} in a very direct way. Namely, for RS, one 
simply defines the $s_n^{\rm RS}$ to equal the asymptotic 
coefficients, that is 
\begin{eqnarray}
\label{eq:snasympRS}
s_{n}^{\rm RS}(\mu/\mu_f) =
N \frac{\mu}{\mu_f}\,\acb_{n}\,.
\end{eqnarray}
While this expression could be used for any $\mu/\mu_f$, the 
implementation proposed in \cite{Pineda:2001zq} first assumes 
$\mu=\mu_f$, in which case
\begin{eqnarray}
 \delta m_{\rm RS}(\mu_f) = \mu_f N \sum_{n=1}^\infty \acb_{n}\,
\alpha_s^{n}(\mu_f)\,,
\label{delmRS}
\end{eqnarray}
which by construction subtracts the leading renormalon divergence 
of the series \eqref{eq:poleMSbar}, and then replaces 
$\alpha_s(\mu_f)$ by its series expansion in $\alpha_s(\mu)$, 
where $\mu$ is the scale at which the pole to $\MSB$ series is 
evaluated.  For the bottom and charm mass, 
the effectiveness of this subtraction is 
analyzed in detail in \cite{Ayala:2014yxa}, which also discusses 
variants of this definition.

A drawback of the RS mass definition is that it needs a precise 
determination of the normalization $N$, which depends on the method 
employed and  further on the number of light flavours, 
see \eqref{norm}. To fully define the RS mass, one needs to 
provide the order to which $\acb_{n}$ is included according to 
\eqref{poleasymp}, and specify the value of $N$.

\subsection{MSR mass}

Another simple realization of the general subtraction 
condition \eqref{eq:snasymp} that avoids the drawback of the 
RS mass definition is to set the $s_n$ equal to $c_n$ and 
simply replace $\mr(\mu_r)$ in \eqref{eq:cnasymp} by the subtraction 
scale $\mu_f$ \cite{Hoang:2008yj,Hoang:2017suc}. More precisely, 
for $\mu=\mu_f$, which is assumed here, we define 
\begin{eqnarray}
\label{eq:snasympMSR}
s_{n}^{\rm MSR}{}_{|\mu=\mu_f} = c_n(\mB,\mB,\mB)\,,
\end{eqnarray}
where the pole to $\MSB$ mass conversion coefficients are 
evaluated at $\mu=\mu_r=\mr(\mu_r)=\mB$, in which case they 
are pure numbers. This gives 
the ``practical version'' \cite{Hoang:2017suc}
of the MSR mass definition 
\begin{eqnarray}
\delta m_{\rm MSR}(\mu_f) = \mu_f \sum_{n=1}^\infty c_{n}(\mB,\mB,\mB)
\,\alpha_s^{n}(\mu_f)\,.
\label{delmMSR}
\end{eqnarray}
The requirement  \eqref{eq:snasymp} is satisfied since 
according to \eqref{eq:cnasymp}
\begin{eqnarray}
s_{n}^{\rm MSR}{}_{|\mu=\mu_f} = c_n(\mB,\mB,\mB) 
 & \underset{n\to\infty}\longrightarrow &  N \,\acb_{n}\,.
\end{eqnarray}
The MSR mass subtraction is straightforward to implement, once 
the pole to $\MSB$ mass conversion coefficients are given. 
The MSR mass interpolates between $\mB$ for $\mu_f=\mB$ and 
the pole mass for $\mu_f=0$, although the latter limit cannot 
be taken as the coupling $\alpha_s(\mu_f)$ flows into the 
strong-coupling regime. The efficiency of the MSR mass 
subtraction is analyzed in detail in \cite{Hoang:2017suc}.

Both the RS and MSR mass satisfy a simple renormalization 
group equation in the subtraction scale, as discussed above. 
As for $\mu_f=\mB$, the MSR mass equals the $\MSB$ mass 
$\mB$, the relation between $m_{\rm MSR}(\mu_f)$ and 
$\mB$ can be obtained conveniently by solving the RGE 
equation, see \cite{Hoang:2017suc}. Alternatively, as for the 
RS scheme, one can replace 
$\alpha_s(\mu_f)$ in \eqref{delmMSR} by its series expansion in 
$\alpha_s(\mu)$, where $\mu$ is the scale at which the pole to 
$\MSB$ series is evaluated, as long as $\ln(\mu/\mu_f)$ is 
small enough not to require resummation. 

\subsection{PS mass}
\label{sec:PS}

The potential-subtracted (PS) 
mass \cite{Beneke:1998rk} is the first of two renormalon-free, 
short-distance, on-shell masses, which are motivated and defined in 
terms of another physical quantity than the pole mass. A non-relativistic 
system of heavy quark and anti-quark in a colour-singlet configuration 
experiences an attractive potential force, whose leading term 
is the Coulomb potential. In momentum space,  
\begin{equation}
\label{coulomb}
\tilde{V}(\bff{q}) = -\frac{4\pi C_F}{\bff{q}^{\,2}}\,
v_c(\alpha_s(\mu),q/\mu)\,,
\end{equation}
where $q=|\bff{q}|$ and $v_c=\alpha_s+\ldots$ incorporates the 
loop corrections to the tree-level potential.  

The PS scheme is based on the observation that there is a cancellation 
of the leading divergent series behaviour in the combination 
$2 m+[V(r)]_{\rm Coulomb}$. This 
can be seen explicitly at the one-loop order and in the large-$\beta_0$ 
approximation \cite{Beneke:1998rk,Hoang:1998nz,PinedaPhD}, and by a 
diagrammatic argument at two loops \cite{Beneke:1998rk} and beyond. 
The cancellation expresses the fact that while the separation of 
the total energy of the quarkonium-like system into the quark pole masses 
and binding energy is ambiguous (as was the case for $m+\bar{\Lambda}$ 
for a heavy-light system), the total energy is physical and 
unambiguous. The PS mass at subtraction scale $\mu_f$ is defined by 
\begin{eqnarray}
\label{PSdef}
m_{\rm PS}(\mu_f) &=& m+\frac{1}{2}\!
\int\limits_{|\mbox{\scriptsize $\bff{q}$}\,|<\mu_f}
\!\!\!\!\frac{d^3 \bff{q}}{(2\pi)^3}\,\tilde{V}(\bff{q})\,, 
\end{eqnarray}
which removes the leading IR contributions to the self-energy from 
$q=|\bff{q}|<\mu_f$.

The series expansion of $v_c(\alpha_s(\mu),q/\mu)$ appearing in 
the Coulomb 
potential \eqref{coulomb} is conventionally written in the form
\begin{equation}
\label{vc}
v_c(\alpha_s(\mu),q/\mu){}_{|\mu=q} = \alpha_s(q)+
\sum_{n=1}^\infty a_n\left(\frac{\alpha_s(q)}{4\pi}\right)^{\!n+1}
+\left(\frac{\alpha_s(q)}{4\pi}\right)^{3} 8\pi^2 C_A^3 
\ln\frac{\nu^2}{\bff{q}^2} + \ldots
\end{equation} 
The coefficients  $a_{1,2,3}$ are known, where $a_3$ 
refers to the  three-loop colour-singlet Coulomb potential 
\cite{Anzai:2009tm,Smirnov:2009fh,Lee:2016cgz}. I note that the 
potential here is not defined in terms of a Wilson loop, but 
as a matching coefficient \cite{Beneke:1998jj} to 
potential non-relativistic QCD (PNRQCD)
\cite{Pineda:1997bj,Beneke:1999qg}, defined with minimal 
subtraction. The last term in \eqref{vc} is the first of an infinite 
series of terms, which contains an explicit  
dependence on the factorization or PNRQCD matching 
scale $\nu$ (to be distinguished from 
$\mu$), which arises 
from an IR divergence related to the ultrasoft scale. 
 
Up to the third order to which the potential is currently known, 
performing the integration over $\bff{q}$ yields 
\cite{Beneke:1998rk,Beneke:2005hg}\footnote{In \cite{Beneke:2005hg}
$\nu=\mu_f$ was assumed.}
\begin{eqnarray}
\delta m_{\rm PS}(\mu_f)
&=&
-\frac{1}{2}\int_{q \leq \mu_f}
 \frac{d^3 {\bff{q}}}{(2\pi)^3}\,
 \tilde{V}(\bff{q}) 
\nonumber \\
&=&
\frac{\mu_f C_F\alpha_s(\mu)}{\pi}
\Bigg[1 
+ \frac{\alpha_s(\mu)}{4\pi} \Big(2 \beta_0\,l_1 +a_1\Big) 
\nonumber\\[0cm]
&& + \,\bigg(\frac{\alpha_s(\mu)}{4\pi}\bigg)^{\!2} 
     \bigg( 4 \beta_0^2\,l_2+2\,\Big(2 a_1\beta_0+\beta_1\Big )l_1
     + a_2 \bigg) 
\nonumber \\
&& 
+ \,\bigg(\frac{\alpha_s(\mu)}{4\pi}\bigg)^{\!3} 
  \bigg(   8 \beta_0^3 l_3 + 
   4 \Big(3 a_1\beta_0^2+\frac{5}{2}\beta_0\beta_1\Big) l_2 
   +2 \Big(3 a_2\beta_0+2 a_1\beta_1+\beta_2\Big) l_1 
\nonumber \\
&& \hspace{1.8cm}    
+ \,a_3 + 
   16 \pi^2 C_A^3 \left[\ln\frac{\nu}{\mu_f}+1\right]\bigg)
\Bigg],
 \label{PSmass}
\end{eqnarray}
where 
\begin{eqnarray}
l_1&=&\ln(\mu/\mu_f)+1\,,
\nonumber\\
l_2&=&\ln^2(\mu/\mu_f)+2\ln(\mu/\mu_f)+2\,,
\nonumber\\
l_3&=&\ln^3(\mu/\mu_f)+3 \ln^2(\mu/\mu_f)+6\ln(\mu/\mu_f)+6\,.
\end{eqnarray}
To fully specify the PS mass definition, a value of $\nu$ must be chosen 
and the standard value is $\nu=\mu_f$, which sets the logarithm 
$\ln(\nu/\mu_f)$ to zero in the last line. This still leaves 
a constant term $16\pi^2 C_A^3 =4263.67...$, which is large 
compared to $a_3=1461.32...$ (quoted for $n_l=5$). Since the 
former is related to an ultrasoft rather than potential effect, another 
well-motivated choice is $\nu=\mu_f e^{-1}$, which nullifies the 
entire square bracket in the last line, resp. the extra term 
in \eqref{vc} at this accuracy. I will refer to this choice 
as the PS$^*$ definition. The difference between the PS and PS$^*$ mass 
is largely irrelevant when the expanded formula \eqref{PSmass} is 
used, since the $n$th order term is dominated by the $l_i$ terms, 
originating from the running coupling in lower order terms,
for relevant values of $\mu_f$. On the other hand, the difference 
affects directly the size of the last presently known term in the 
anomalous dimension and RGE below.

It is instructive 
to interpret $v_c(q)=v_c(\alpha_s(\mu),q/\mu){}_{|\mu=q}=\alpha_s(q)
+\ldots$ as an effective coupling, and write the mass subtraction term 
as 
\begin{equation}
\delta m_{\rm PS}(\mu_f) = 
\frac{C_F}{\pi} \int_0^{\mu_f} dq\,v_c(q)\,. 
\label{delMSfromvc}
\end{equation}
The $\mu_f$ evolution of the PS mass is governed by the anomalous 
dimension
\begin{equation}
\gamma_{\rm PS}(\alpha_s(\mu_f)) = - \frac{d m_{\rm PS}(\mu_f)}{d\mu_f}
= \frac{C_F}{\pi} v_c(\mu_f)\,.
\end{equation}
The solution is evidently (see \eqref{delMSfromvc})
\begin{equation}
\Big[\delta m_{\rm PS}(\mu_f)\Big]^{\mu_{f2}}_{\mu_{f1}} 
= \frac{C_F}{\pi} \int^{\mu_{f2}}_{\mu_{f1}} dq\,v_c(q)\,,
\end{equation}
which allows us to compute $m_{\rm PS}(\mu_f)$ at widely different 
scales $\mu_{f1},\mu_{f2}$ without large 
logarithms by integrating the expansion of $v_c(q)$ 
in terms of the running $\MSB$ coupling $\alpha_s(q)$. The 
leading-logarithmic solution can be expressed in terms of the 
exponential-integral function.

The PS scheme is especially well-suited for quarkonium-like 
systems including open $Q \bar Q$ systems near threshold, since it 
subtracts the leading IR contributions explicitly 
already in low orders of perturbation theory. Prime examples 
are quarkonium masses at next-to-next-to-next-to-leading order 
(NNNLO)~\cite{Beneke:2005hg}, the determination of the 
bottom-quark mass from high moments of the pair production cross 
section at second \cite{Beneke:1999fe} and third order 
\cite{Beneke:2014pta} in PNRQCD, and in particular precision 
calculations of top-quark pair production near threshold
to NNLO and NNNLO \cite{Beneke:1999qg,Beneke:2015kwa} in the 
PS scheme. For $Q\bar{Q}$ systems near threshold, 
the scale $\mu_f$ should be chosen parametrically 
of order $m v\sim m \alpha_s$ 
such that $\delta m_{\rm PS} \sim m v^2$, 
in order not to violate the power counting of the  
non-relativistic expansion. With this choice, the 
relation (\ref{PSdef}) is already accurate to order 
$m \alpha_s^5$.

\subsection{Kinetic mass}

The kinetic mass scheme is another physical scheme, in this case 
related to the physics of semi-leptonic decays of 
heavy-light mesons \cite{Bigi:1994ga,Bigi:1996si}. 
The pseudoscalar $B$ meson mass has the heavy-quark 
expansion (cf. \eqref{mesonmass})
\begin{equation}
m_B = m_b+\bar{\Lambda}+\frac{\mu_\pi^2-\mu_G^2}{2 m_b}+\ldots,
\end{equation}
where $\mu_\pi^2$ and $\mu_G^2$ are the $B$-meson matrix elements 
of the kinetic energy and chromo-magnetic operators, respectively. The 
kinetic mass can be understood as a perturbative evaluation of this formula, 
in which the matrix elements include loop momentum integration 
regions below the scale $\mu_f$: 
\begin{eqnarray}
m_{\rm kin}(\mu_f) &=& m \underbrace{- [\bar{\Lambda}(\mu_f)]_{\rm pert}
- \left[\frac{\mu_\pi^2(\mu_f)}{2 m_{\rm kin}(\mu_f)}\right]_{\rm pert}}_{-\delta m_{\rm kin}(\mu_f)}
+\ldots.
\label{mkindef}
\end{eqnarray}
The matrix elements on the right-hand side subtract the long-distance 
sensitive contributions to the pole mass order by order in 
$\mu_f/m_Q$ and $\alpha_s$. Comparing to \eqref{mesonmassmod}, we 
note that the kinetic mass definition not only subtracts the leading 
IR renormalon divergence through $ [\bar{\Lambda}(\mu_f)]_{\rm pert}$, 
but also the IR sensitivity at subleading power $\lambdaQCD^2/m$.

Up to now the discussion has been general. The kinetic scheme is defined 
by providing a concrete prescription for calculating the perturbative 
matrix elements in terms of the perturbative evaluation of short-distance 
observables. It relies on the fact that the $\bar\Lambda$ parameter and 
kinetic-energy matrix element appear in the heavy-quark expansion 
of the dilepton-differential spectrum of inclusive semi-leptonic 
$B\to X_c\ell\bar{\nu}$ decays \cite{Manohar:1993qn,Blok:1993va}, which can be 
constructed from the imaginary part of the two-point function 
of the $b\to c$ transition current. The convention for the kinetic mass 
used in the literature employs an indirect definition of the 
matrix elements through heavy flavour sum 
rules \cite{Bigi:1994ga,Bigi:1996si} in the small-velocity limit, 
which is rather complicated when 
compared with the other three mass definitions above. The central 
quantity is the forward amplitude 
\begin{eqnarray}
 T(q) &=& \frac{i}{2m_Q} \int {\rm d}^4{x \,} e^{-iqx}
\langle Q|T J(x)J^\dagger(0)|Q\rangle
\label{Tforward}
\end{eqnarray}
and its discontinuity
\begin{eqnarray}
W(q) &=& 2 \,\mbox{Im}\,[ T(q) ]\,.
\end{eqnarray}
Since the perturbative matrix elements (\ref{mkindef}) to be computed 
are spin-independent and universal, instead of the physical 
V-A current, one can define the kinetic scheme by adopting  
the scalar current $J=Q Q^\prime$ 
provided $Q$ and $Q^\prime$ are heavy. To simplify further, 
one can set $Q^\prime = Q$ and compute the forward scattering 
of the heavy quark off the current $J$ with momentum $q=(q^0,\bff{q})$ 
in the limit when $\bff{v}=\bff{q}/m_Q\ll 1$. To isolate the 
IR region of the final state momenta, the total energy of the 
quark and gluon final state $X$ excluding the heavy quark $Q$ 
is restricted to 
\begin{equation}
\omega \equiv q^0-\Big[\sqrt{m_Q^2+\bff{q}^2}-m_Q\Big] <\mu_f\,.
\end{equation} 
Employing the variables $(\omega,\bff{v})$ instead of 
$(q^0,\bff{q})$, the mass subtractions are defined in terms of 
the double limit of the first moments of the spectral function:
\begin{eqnarray}
[\bar{\Lambda}(\mu_f)]_{\rm pert} &=& 
\lim_{\bff{v}\to0}\lim_{m_Q\to\infty} \frac{2}{\bff{v}\,^2} 
      \frac{\displaystyle
      \int_0^{\mu_f} d\omega\,\omega \, W(\omega,\bff{v})}
      {\displaystyle
      \int_0^{\mu_f} d\omega \,W(\omega,\bff{v})}\,,
\\[0.2cm]
[\mu_\pi^2(\mu)]_{\rm pert} &=& 
\lim_{\bff{v}\to0}\lim_{m_b\to\infty} \frac{3}{\bff{v}\,^2} 
 \frac{\displaystyle\int_0^{\mu_f} d\omega\,\omega^2 \, W(\omega,\bff{v})}
      {\displaystyle\int_0^{\mu_f} d\omega \,W(\omega,\bff{v})}\,.
\label{mkinMEs}
\end{eqnarray}
The demonstration that the right-hand sides of these equations 
can be identified with the subtracted $\bar{\Lambda}$ and 
kinetic-energy parameters, which appear in the heavy-quark 
expansion of the meson mass, is given in \cite{Bigi:1996si}.
It is sufficient to compute $W(\omega,\bff{v})$ in an expansion 
in $\omega, |\bff{q}| \ll m_Q$ to order 
\begin{equation}
W(\omega,\bff{v}) = W_{\rm virt}(\bff{v}) \delta(\omega) 
+ \frac{\bff{v}^2}{\omega} W_{\rm real} \Theta(\omega) 
+\mathcal{O}(\omega^0,\bff{v}^4)\,.
\end{equation} 
The two-loop computation \cite{Czarnecki:1997sz} has been known 
for some time, but the three-loop result has been 
obtained only recently \cite{Fael:2020iea,Fael:2020njb}. 
Different from the other three mass schemes, the 
$\mathcal{O}(\alpha_s^4)$ term of $\delta m_{\rm kin}(\mu_f)$ 
is presently not available.

The kinetic scheme is especially well-suited for observables 
derived from semi-leptonic decays of heavy quarks, since it 
subtracts the leading IR contributions explicitly 
already in low orders of perturbation theory. 
Initially, it was suggested to eliminate the leading renormalon 
divergence from the semi-leptonic width by replacing the pole 
mass by the $\MSB$ mass \cite{Ball:1995wa}, but this does not 
improve the behaviour in low orders. The comparison to 
the series convergence when renormalon-free on-shell masses 
are used (Table~2 of \cite{Beneke:1999zr}) clearly shows the 
latter's advantage. The kinetic scheme was used for bottom quark 
mass and $|V_{cb}|$ determinations at second 
order~\cite{Gambino:2013rza}. Recently, the semi-leptonic 
decay width was calculated to NNNLO and the effectiveness 
of the kinetic mass scheme was demonstrated for the inclusive 
rate for the first time at this order \cite{Fael:2020tow}. 

\subsection{Comparison}

\begin{table}
\vskip0.5cm
\caption{Comparison of top quark mass definitions  
for $\mu_f=20\,$GeV for given $\MSB$ mass 
$\mB_t$. $\alpha_s(m_Z)=(0.1180\pm 0.0010)$. ``$n$-loop'' refers to the 
value of the $n$-loop contribution to the mass. All numbers in GeV.}
\vskip0.2cm
\label{tabmXtop}  
\setlength{\extrarowheight}{0.14cm}
\centering
\begin{tabular}{|c|c|c|c|c|c|c|}
\hline 
&&&&&&\\[-0.4cm]
 top & $\mB_t$ & 1-loop & 2-loop &
3-loop & 4-loop & Sum\\[0.2cm]
\hline
$m_t$ & 163.643 & 7.531 & 1.606 & 0.494 & 0.194 & $173.468^{+0.101}_{-0.101}$ \\
\hline
$m_{t,\rm RS}$ & 163.643& 6.123 & 1.079 & 0.239 & 0.048& $171.132^{+0.071}_{-0.071}$ \\
\hline
$m_{t,\rm MSR}$ & 163.643& 6.611 & 1.154 & 0.242 & 0.042 & $171.691^{+0.076}_{-0.076}$\\
\hline
$m_{t,\rm PS}$ & 163.643&  6.611 & 1.190 & 0.265 & 0.048& $171.756^{+0.077}_{-0.077}$  \\
\hline
$m_{t,\rm PS^*}$ & 163.643&  6.611 & 1.190 & 0.265 & 0.051& $171.759^{+0.078}_{-0.077}$  \\
\hline
$m_{t,\rm kin}$ & 163.643 & 6.248& 0.995& 0.180 &--- & $171.065^{+0.068}_{-0.068}$ \\
\hline
\end{tabular}
\vskip1.5cm
\caption{Comparison of bottom quark mass definitions 
for $\mu_f=2\,$GeV for given $\MSB$ mass 
$\mB_b$. $\alpha_s(m_Z)=(0.1180\pm 0.0010)$. ``$n$-loop'' refers to the 
value of the $n$-loop contribution to the mass. All numbers in GeV.}
\vskip0.2cm
\label{tabmXbottom}  
\setlength{\extrarowheight}{0.14cm}
\centering
\begin{tabular}{|c|c|c|c|c|c|c|}
\hline 
&&&&&&\\[-0.4cm]
bottom & $\mB_b$ & 1-loop & 2-loop &
3-loop & 4-loop & Sum\\[0.2cm]
\hline
$m_b$ & 4.200 & 0.400 & 0.199 & 0.145 & 0.135 & $5.079^{+0.031}_{-0.029}$ \\
\hline
$m_{b,\rm RS}$ & 4.200 & 0.126 & 0.065& 0.028& 0.007& $4.425^{+0.006}_{-0.006}$\\
\hline
$m_{b,\rm MSR}$ & 4.200 & 0.210  & 0.062 & 0.020 & 0.001 & $4.493^{+0.007}_{-0.007}$\\
\hline
$m_{b,\rm PS}$ & 4.200 & 0.210 & 0.080 & 0.032 & 0.000 & $4.521^{+0.008}_{-0.008}$\\
\hline
$m_{b,\rm PS^*}$ & 4.200 & 0.210 & 0.080 & 0.032 & 0.005 & $4.526^{+0.008}_{-0.008}$ \\
\hline
$m_{b,\rm kin}$ & 4.200 & 0.101 & 0.004 & -0.002 & --- & $4.303^{+0.002}_{-0.002}$ \\
\hline
\end{tabular}
\vskip1.5cm
\caption{Comparison of charm quark mass definitions 
for $\mu_f=1\,$GeV for given $\MSB$ mass 
$\mB_c$. $\alpha_s(m_Z)=(0.1180\pm 0.0010)$. ``$n$-loop'' refers to the 
value of the $n$-loop contribution to the mass. All numbers in GeV.}
\vskip0.2cm
\label{tabmXcharm}  
\setlength{\extrarowheight}{0.14cm}
\centering
\begin{tabular}{|c|c|c|c|c|c|c|}
\hline 
&&&&&&\\[-0.4cm]
charm & $\mB_c$ & 1-loop & 2-loop &
3-loop & 4-loop & Sum\\[0.2cm]
\hline
$m_c$ & 1.280 & 0.211 & 0.202 & 0.282 & 0.510 & $2.486^{+0.126}_{-0.109}$ \\
\hline
$m_{c,\rm RS}$ & 1.280 & -0.017 &0.037 & 0.026& 0.005 & $1.331^{+0.006}_{-0.005}$\\
\hline
$m_{c,\rm MSR}$ & 1.280 & 0.046 & 0.022 & 0.010& -0.002& $1.356^{+0.004}_{-0.004}$\\
\hline
$m_{c,\rm PS}$ & 1.280 &0.046  & 0.052 & 0.034 & -0.019 & $1.393^{+0.006}_{-0.006}$ \\
\hline
$m_{c,\rm PS^*}$ & 1.280 & 0.046  & 0.052 & 0.034 & 0.002 & $1.414^{+0.009}_{-0.008}$ \\
\hline
$m_{c,\rm kin}$ & 1.280 & -0.073 & -0.062 & -0.017 & --- & $1.128^{+0.008}_{-0.009}$\\
\hline
\end{tabular}
\end{table}

Once any of the leading renormalon-free, on-shell, short-distance 
masses has been determined from some observable, one is eventually 
interested in converting them to the $\MSB$ reference mass $\mB$. 
Over the past twenty years the accuracy of the mass definitions 
and observables has improved by one order (typically from two-loop 
to three-loop, and three-loop to four-loop for the pole to 
$\MSB$ mass series). It is therefore timely to update and extend the 
comparison  \cite{Beneke:1999zr} of different definitions (see 
also \cite{Marquard:2016dcn}). 

The purpose of the following comparison is to display the good 
behaviour of the relation between the various subtracted masses 
and the $\MSB$ mass contrary to the pole mass series 
\eqref{mtseries} -- \eqref{mcseries}. As above, the \MSB~masses 
are fixed to $\mB_t=163.643\,$GeV, $\mB_b=4.20\,$GeV and 
$\mB_c=1.28\,$GeV. The strong coupling is taken to be 
$\alpha_s^{(5)}(m_Z)=0.1180\pm 0.0010$ at the scale 
$m_Z=91.1876~$GeV, and the series coefficients 
are evaluated at $\mu=\mu_m=\mB$ in an expansion 
in $\alpha_s^{(5)}(\mB_t)=0.1084$, $\alpha_s^{(4)}(\mB_b)=0.2246$, 
$\alpha_s^{(3)}(\mB_c)=0.3889$, for top, bottom and charm, 
respectively. The subtraction scale $\mu_f$ is chosen 
to be $\mu_f=(20,2,1)\,$GeV for top, bottom and charm, 
equal for all mass definitions for the sake of 
comparison, even if different ``canonical'' values are often
adopted for the different schemes. Internal mass effects 
are neglected, since they are not always known with 
comparable precision. In case of mass schemes originally defined 
in terms of $\alpha_s(\mu_f)$, the series have been converted into 
expansions in $\alpha_s(\mB_Q)$ with the required four-loop 
accuracy. 

I use private code for the RS, MSR and PS mass. The RS 
and PS scheme (for $\nu=\mu_f$) is also implemented in 
{\tt CRunDec3.1} \cite{Herren:2017osy} and the PS mass also in 
$\tt{QQbar\_threshold}$ \cite{Beneke:2016kkb}. In case of the 
RS scheme I adapted the normalization constants $N_m = 0.563$ 
($n_l=3$), $N_m = 0.547$ ($n_l=4$), 
$N_m = 0.527$ ($n_l=3$) given in \cite{Ayala:2014yxa} (and hard-coded 
in {\tt CRunDec3.1}) to the values from \eqref{norm}. Finally, 
the kinetic mass $m_{\rm kin}$ is implemented for massless internal 
quarks with  {\tt CRunDec3.1}'s 
{\tt{mMS2mKIN[}$\mB_Q$, {0, 0}, "[as]"*$\alpha_s^{(n_l)}(\mB_Q)$, 
$\mB_Q$, $\mu_f$, $n_l$, $n_l$, 3, ""]} call, presently available only to 
three-loop accuracy.

The results are summarized in Tables~\ref{tabmXtop} 
to~\ref{tabmXcharm}. It is evident that in terms of the size of 
mass corrections up to the shown four-loop order, all mass schemes are 
rather similar with exception of the kinetic scheme. In all 
cases, one observes a spectacular improvement of convergence 
relative to the pole mass series given in the first line.
One expects the cancellation of the leading renormalon to 
become more and more effective in higher orders measured relative to the order of the minimal term, which can be 
seen explicitly by comparing the top, bottom and charm tables. 
The effect is particularly dramatic for the charm mass, for 
which the pole mass series starts diverging beyond the two-loop 
order, while the renormalon-subtracted masses are still 
well-behaved at the fourth order, with corrections in the 
few MeV range. The top pole mass series is still in the regime 
of decreasing coefficients at the four-loop order, albeit 
slowly, hence the relative improvement of the subtraction should 
increase in the next orders. The four-loop coefficient is 
typically $(40-50)\,$MeV, but the size of the next unknown term
can be assumed to be small enough relative to the experimental 
precision that can be attained in the future, even from 
the scan of the pair production cross section in $e^+e^-$ 
collisions \cite{Simon:2019axh}. 

``Sum'' in the last column of the tables refers to the sum of 
the terms up to the four-loop 
order shown, and the error attached quantifies the variation 
of the sum under a variation of $\alpha_s(m_Z)$ by $\pm 0.001$.
It is apparent that once leading renormalon-free, on-shell masses
are employed, the limitation of the accuracy of their relations 
to the $ \MSB$ mass is (currently) no longer determined by the 
convergence of the expansion, but by the precision of 
$\alpha_s(m_Z)$. For the case of the top quark, this uncertainty 
is mainly caused by the large one-loop correction of a few GeV 
to be compared to the ultimate precision to which the subtracted masses 
can be obtained theoretically and (in principle) experimentally. 
Unlike the bottom and charm masses, to make use of this precision 
requires better knowledge of the strong coupling.

\paragraph{Acknowledgement.} I thank F. Herren and M. Steinhauser for 
helpful communication on {\tt RunDec}'s implementation of the kinetic 
scheme.

\bibliographystyle{JHEP}

\providecommand{\href}[2]{#2}\begingroup\raggedright\endgroup


\end{document}